\documentclass[apsrev, preprint, showpacs, showkeys]{revtex4-1}
\usepackage{amsmath}
\usepackage{amssymb}
\usepackage{graphicx}
\usepackage{hyperref}
\usepackage{subfigure}
\usepackage{setspace}

\usepackage{color}

\begin{document}

\title{ Population Genetics with Fluctuating Population Sizes }
\author{Thiparat Chotibut}
\email{Electronic address: thiparatc@gmail.com}
\author{David R. Nelson}
\email{Electronic address: nelson@physics.harvard.edu}
\affiliation{Department of Physics, Harvard University, Cambridge, Massachusetts
02138, USA}

\date{\today}
\begin{abstract}
Standard neutral  population genetics theory with a strictly
fixed population size has important limitations. An alternative model that
allows independently fluctuating population sizes and reproduces  the standard
neutral evolution is reviewed. We then study a situation such that the
competing species are neutral at the equilibrium
population size but population size fluctuations nevertheless favor fixation
of
one species over the other. In this case, a  separation of timescales
emerges naturally and allows  adiabatic elimination of a fast population
size variable to deduce the fluctuations-induced selection dynamics near
the equilibrium population size. The results highlight the incompleteness
of the
standard population genetics with a strictly fixed population size. 
\end{abstract}
\maketitle
\section{Introduction}
\label{sec:intro} 

Evolutionary processes are ubiquitous in living systems. Organisms reproduce
and pass on their genes to their descendants. Depending on environmental
conditions
and interactions among organisms in living populations,  fitter organisms
tend to reproduce faster by \textit{natural} \textit{selection}. However,
fitter organisms in a particular generation may also give birth to fewer
descendants  by random chance;
 these noisy statistical fluctuations in the reproduction rates are termed
\textit{genetic
drift} 
 in population genetics \cite{Gillespie:2010uq,Ewens:2004kx}. In well-mixed
competition experiments, in which different species
of microbes grow in a vigorously shaken test tube or in a chemostat,
both selection and genetic drift influence the evolutionary dynamics that
determines the genetic
composition of  populations \cite{Elena:2003sf,Desai:2013jp,Barrick:2013aa}.
In this work, we neglect  the less frequent changes due to additional spontaneous
mutations, and focus as well on the dynamics of asexual organisms. 

Although advances in experimental evolution have revealed the interplay
between evolutionary dynamics  and the dynamics of population size \cite{Barrick:2013aa,Dai:2012wd,Sanchez:2013eu,Griffin:2004qe},
standard theoretical  frameworks are often limited to evolutionary dynamics
in a strictly fixed population size \cite{Ewens:2004kx,Nowak:2006uq,Gillespie:2010uq,Hartl:1997hb}.
Several population genetics works  addressed how evolutionary dynamics  is
affected by  a deterministically
changing population size,
such as during exponential, logistic, or cyclic  population growth  \cite{Otto723},
and during the population bottlenecks inherent in serial transfer experiments
 \cite{Wahl2002,Wahl:2001aa} (for a review, see Ref.  \cite{Patwa:2008aa}.)
In this paper, we study instead evolutionary
dynamics with a  stochastically fluctuating population size  and show that,
 even if
 two competing species
are neutral at the equilibrium population size,  the coupling between evolutionary
dynamics and the dynamics of population size can lead to a  fluctuation-induced
selection mechanism that favors one species over the other, in a way that
is inherited from the dynamics away from the equilibrium population size.
Although the  model we study    is mathematically similar to those in
 Refs. \cite{Parsons:2008aa,Parsons:2007aa, Lin:2012wf, Kogan:2014aa}, we exploit here the tools of statistical physics, which leads via Sec. \ref{sec: EffectiveEvo} to a number of novel predictions tabulated in
Sec. \ref{subsec: 4C(dimensional_red)}.

We first
review the  standard model
of \textit{neutral} evolution in this introduction section. For two competing
neutral species in a well-mixed environment,  both species
grow  on average at the same rate and the genetic compositions  remain unchanged
in the limit of infinitely large population size. However, in finite populations,
genetic compositions can also be influenced by
fluctuating evolutionary forces from random birth and death events.  Random
fluctuations in the reproductive rate, or genetic drift,
is one of the central concepts  in population genetics embodied in
the foundational work of Fisher \cite{Fisher:1930aa} and Wright
\cite{Wright:1931aa}. In the Wright-Fisher model consisting of $N$ neutral
haploid individuals, random members of the   parental
generation are  chosen to give birth via cell division to a
generation of daughters. The generations
are
assumed non-overlapping, and the random sampling process culls the offspring
to insure that  the
daughter generation still contains exactly $N$ individuals. This random sampling
with replacement
 simulates random birth and death events of the parental generation. For
two
neutral species $1$ and $2$, this process generates fluctuations in the species
frequency  (relative fraction) as follows:   let $X_t $ denote the number
of species
$1$ in generation $t$;  the conditional probability that $X_{t+1}=n$ given
$X_{t} = m$ is the binomial distribution:
\begin{equation}
P(X_{t+1}=n|X_t = m)=  {N \choose n}\left( \frac{m}{N} \right)^n\left(1-
\frac{m}{N} \right)^{N-n}.
\label{eqn: chp1_binomialsamp}
\end{equation}
Using properties of the binomial distribution, one finds that the mean species
frequency
remains
unchanged.
However, the species frequency $f_t \equiv X_t/N$ fluctuates with a variance
that depends on both the population size $N$ and the species frequency of
the
parental generation $f$  as 
\begin{equation}
\text{Var}(f_{t+1}|f_t = f) = \frac{f(1-f)}{N}.
\label{eqn: chp1_binomial_gendrift}
\end{equation}
\begin{figure}
\includegraphics[width=\textwidth]{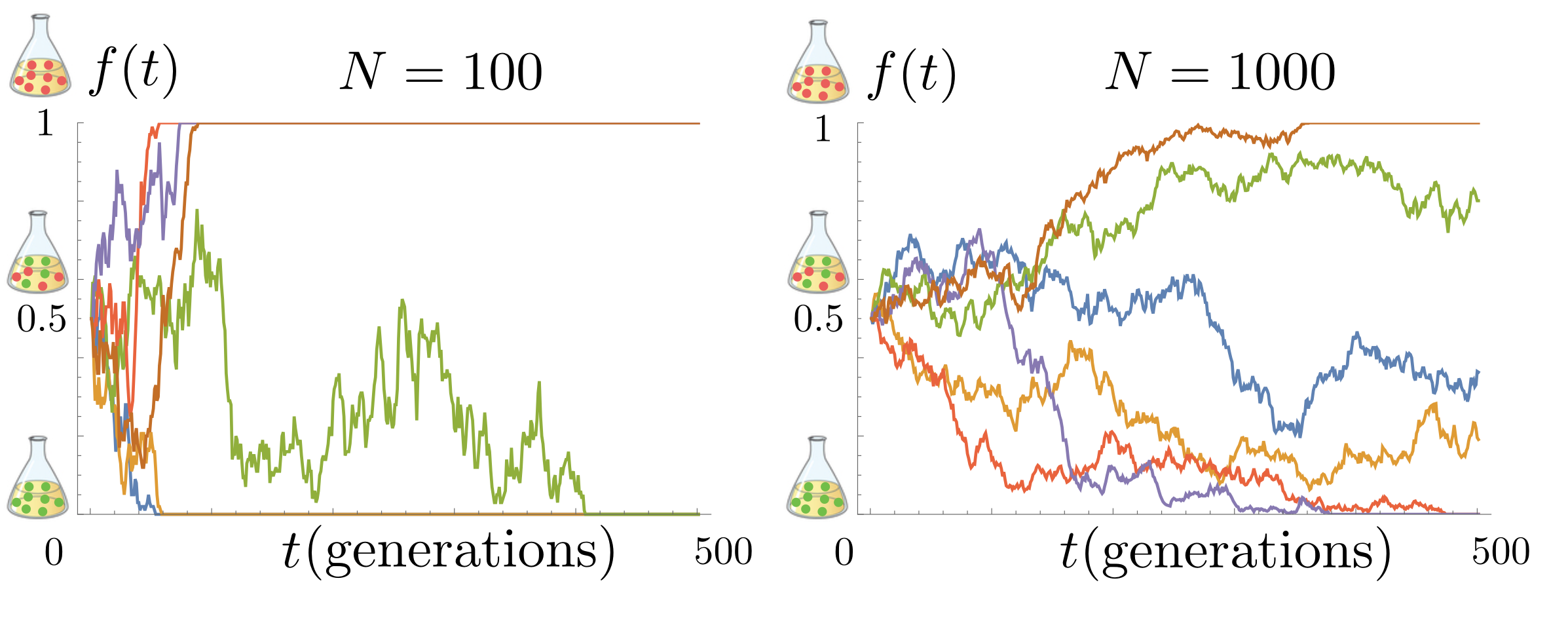}
\caption{ (Color Online) Unbiased random walk  in the allele frequency $f(t)$ due to genetic
drift in neutral haploid asexual populations  with the population size  
$N=100$ (left) and $N=1000$ (right). Each color of fluctuating paths represents
different realizations of genetic drift simulated from the 
Wright-Fisher model. Initially, both populations consist of an equal mixture
of \ two alleles (i.e. variants);  however, microscopic fluctuations due
to genetic drift
 lead to an eventual fixation, an irreversible macroscopic change in the
compositions. The standard deviation of these fluctuations 
scales as $1/\sqrt{N},$ which, from the result of the first passage time
of
an unbiased random walk \cite{Redner:2001aa}, implies the mean time to fixation
scales as $N$.}
\label{fig: ch1_wright-fisher}
\end{figure}
 This discrete-time unbiased random walk in the genetic composition
of populations exemplifies genetic drift, whose effect becomes more pronounced
at smaller population size. Despite being neutral (identical reproduction
rates on average), one of the species can take over the populations (\textit{fixation})
by chance. Fig. \ref{fig: ch1_wright-fisher} illustrates genetic drift for
the Wright-Fisher model.

A variant of the Wright-Fisher model for genetic drift is the Moran model,
which does not assume non-overlapping generations. Although the original
model is formulated in discrete time \cite{Moran:1958aa,Moran:1962pi}, we
introduce the continuous time version here as it is more relevant to
 statistical physics in the context of the Master equation. Recall that,
in a continuous-time discrete-state Markov process, the time evolution
of the probability distribution $P(\boldsymbol n, t)$ for finding the system
in a discrete state $\boldsymbol n$ at time $t$ evolves according to the
Master equation \cite{Van-Kampen:1992tk}:
\begin{equation}
\partial_tP(\boldsymbol n, t) = \sum_{\boldsymbol n'}\left[ W(\boldsymbol
n|\boldsymbol n')P(\boldsymbol n',t) -  W(\boldsymbol
n'|\boldsymbol n)P(\boldsymbol n,t)\right],
\label{eqn: ch1_master_eq} 
\end{equation}
where $W(\boldsymbol n| \boldsymbol n')$ is the transition rate from the
configuration $\boldsymbol n'$ to $\boldsymbol n.$ The Moran model is a Markov
process that specifies the transition rate by a continuous-time sampling
with replacement.
In a finite population of size $N$ with   $n$ representatives of species
$1$ and $N-n$ of species
$2$,  two individuals  are sampled at a  rate $\mu$; one is chosen to reproduce
and the other is chosen to die to ensure the population size remains constant.
The
transition rates for  reproduction  of species $1$ (death of species $2$)
and for death of  species $1$ (reproduction of  species $2$) are thus given
by, respectively, 
\begin{equation}
W(n+1|n)= \mu \left(1-\frac{n}{N} \right) \left( \frac{n}{N} \right), 
\label{eqn: ch1_moranrate1}   
\end{equation}
\begin{equation}
W(n-1|n)= \mu \left(1-\frac{n}{N}\right) \left( \frac{n}{N} \right).
\label{eqn: ch1_moranrate2}   
\end{equation}
The Master equation describing the dynamics of species $1$ in the Moran model
reads
\begin{align}
\nonumber \partial_tP(n,  t) = &\left[ W(n|n+1)P(n+1,t) + W(n|n-1)P(n-1,t)\right]
\\  &- \left[ W(n+1|n)P(n,t) + W(n-1|n)P(n,t)\right],
\label{eqn: ch1_moran}   
\end{align}
where the transition rates are given by Eqs.(\ref{eqn: ch1_moranrate1}) and
(\ref{eqn: ch1_moranrate2}).    

In the large $N$ limit, we may promote the
species frequency  $f = n/N$ to
a continuous variable and approximate the discrete Master equation (\ref{eqn:
ch1_moran})
by the Fokker-Planck equation in $f$. Two systematic methods for deriving
the Fokker-Planck
approximation to the discrete Master equation  are the Kramers-Moyal expansion
and the Van-Kampen's system size
expansion \cite{Van-Kampen:1992tk,Risken:1984df}. Both methods require that
we  Taylor expand Eq. (\ref{eqn:
ch1_moran}) to $O(1/N^2)$:

\begin{align*}
\partial_t P(f,t) = &\mu\left(f - \frac{1}{N} \right)\left(1-f +  \frac{1}{N}\right)\left(
1-\frac{1}{N} \partial_f + \frac{1}{2N^2} \partial^2_f\right)P(f,t) \\
&+\mu\left(f + \frac{1}{N} \right)\left(1-f -  \frac{1}{N}\right)\left(
1+\frac{1}{N} \partial_f + \frac{1}{2N^2} \partial^2_f\right)P(f,t)\\
&- 2\mu f(1-f)P(f,t) +O(1/N^{3}).
\end{align*}
Upon defining one generation time as $\tau_g = N\mu^{-1}$, which represents
$N$ random sampling events, the final result is a Fokker-Planck equation
for genetic
drift:\begin{equation}
\partial_t P(f,t) = \frac{1}{\tau_g} \partial^2_f \left[\frac{D_g(f)}{N}
P(f,t) \right],
\label{eqn: moran_FP}
\end{equation}
where the frequency-dependent genetic diffusion coefficient is 
\begin{equation}
\frac{D_g(f)}{N} = \frac{f(1-f)}{N},
\label{eqn: ch1_Moran_DiffuCoeff}
\end{equation}
similar to the variance per generation time of Eq. (\ref{eqn: chp1_binomial_gendrift})
in the Wright-Fisher model. The Fokker-Planck equation  (\ref{eqn: moran_FP})
 describes genetic drift as an unbiased random walk
in the frequency space, provided the two competing species are neutral.

By absorbing $\tau_g$ into the unit of time, one obtains a   stochastic differential
equation associated with Eq. (\ref{eqn: moran_FP}) that reveals the
underlying continuous time stochastic dynamics   
\begin{equation} 
\frac{df}{dt}=  \sqrt{\frac{2\mu D_g(f)}{N}}\Gamma(t),
\label{eqn: ito_rep_drift}
\end{equation}
where $\Gamma(t)$ is the Gaussian white-noise with zero mean $\langle \Gamma(t)
\rangle$ and unit variance $\langle \Gamma(t)\Gamma(t')
\rangle = \delta(t-t')$ \cite{Korolev:2010fk,Risken:1984df}.  To recover
the unbiased Fokker-Planck equation (\ref{eqn: moran_FP}) of population genetics,
 where the fluctuations in a given generation are entirely determined by
the
statistics of the preceding generation, the Ito's interpretation
of Eq. (\ref{eqn:
ito_rep_drift}) must be employed \cite{Gardiner:1985qr,Risken:1984df}.
The stochastic differential equation (\ref{eqn: ito_rep_drift}) implies that
once the system reaches either $f=0$ or $f=1$,  the
dynamics completely stop;  the genetic drift, whose strength is proportional
to the diffusion coefficient
$D_g(f)/N = f(1-f)/N,$ vanishes at these states. Since  fluctuations can
drive
the system into  but not away from  $f=0$ and
$f=1$, these  are absorbing states.  

Two important quantities  quantify the fate of the surviving species: 
 \textit{the fixation probability} $u(f)$ and \textit{the
mean fixation time} $\tau(f)$. These are, respectively, the probability that
a species of interest takes over the population and
the average time required for this to happen,
given an initial  composition $f$. Throughout this paper, we shall refer
to fixation as the situation when  species 1 takes over, which is equivalent
to the situation that the system eventually reaches the absorbing state $f=1$.
This first passage problem 
is more conveniently studied in a backward time formulation (backward
Kolmogorov equation) with a target state in mind, rather
than the forward time formulation in the Fokker-Planck equation (forward
Kolmogorov equation) \cite{Risken:1984df,Gardiner:1985qr,Redner:2001aa}.
The backward Kolmogorov equation for the fixation probability and the mean
fixation time associated with Eq. (\ref{eqn: ito_rep_drift}) are, respectively,

\begin{align}
 \frac{D_g(f)}{N}\frac{d^2}{df^2}u_{neutral}(f)&=0, \label{eqn: bke_u} \\
 \frac{D_g(f)}{N}\frac{d^2}{df^2}\tau_{neutral}(f)&=-\frac{1}{\mu},  \label{eqn:
bke_tau}
\end{align}
subject to the boundary conditions $u_{neutral}(0) = 0, u_{neutral}(1) =
1$ and $\tau_{neutral}(0) =
\tau_{neutral}(1) = 0$ \cite{Van-Kampen:1992tk,Gardiner:1985qr}. Integrating
Eqs. (\ref{eqn:
bke_u}) and (\ref{eqn: bke_tau}) yields the standard results \begin{equation}
u_{neutral}(f)=f,  \label{eqn:fixprob_neutral}
\end{equation}
and
\begin{equation}
\tau_{neutral}(f) =-\left(\frac{N}{\mu}\right)\Big[f\ln f+(1-f) \ln (1-f)\Big].\label{eqn:MFT_neutral}
\end{equation}

Although the effect of genetic drift in neutral evolution as embodied in
the
Moran or the Wright-Fisher model are well studied, this framework enforces
 a
\textit{strictly fixed}  population size $N$ through a strictly enforced
growth
condition: the birth of one species necessitates the death of the other,
somewhat like the canonical ensemble in equilibrium statistical mechanics.
In evolution experiments,
 as  well as in natural environments, population size
fluctuations away from a preferred carrying capacity often arise \cite{Barrick:2013aa}.
In Section 2, we discuss a two-species competitive Lotka-Volterra
model that accounts for natural  population growth
and encompasses   neutral evolution at the equilibrium population size. Instead
of artificially enforcing a  strictly fixed population size $N$,  the population
size becomes a dynamical variable $N(t)$ (like the grand canonical ensemble
of statistical mechanics) and couples to the evolutionary
dynamics  $f(t).$ In Section 3, we show that,  while $N(t)$
 fluctuates around a fixed stable equilibrium size $N$,  neutral evolution
with genetic drift of Eq.  (\ref{eqn: ito_rep_drift})
can acquire a fluctuation-induced selection bias as a result of the coupling
between  $f(t)$ and $N(t)$.
Species with a selective disadvantage in the dilute limit
far from the equilibrium population size acquire a selective \textit{advantage}
for
competitions at long times near $N$. After adiabatic
elimination of the fast population size variable, the effective evolutionary
dynamics of \textit{quasi}-neutral
evolution near the equilibrium population size $N$ is determined, and  the
classical population genetics results of Eqs.
(\ref{eqn:fixprob_neutral}) and (\ref{eqn:MFT_neutral})
 are modified.

\section{Neutral Evolution from a Competitive Lotka-Volterra Model}

The two-species competitive Lotka-Volterra model assumes that each species
$S_i$
grows under dilute conditions with rates 
\begin{equation}
 S_{i }\xrightarrow{\mu_i} \label{eqn:Birth}
S_i+S_i, 
\end{equation}
 and competes for limited resources under  crowded conditions with rates
\begin{equation}
S_i + S_j \xrightarrow{\lambda_{ij}} S_j.\label{eqn:Death}
\end{equation}  In an infinitely large population  and in the
absence of interspecies competition ($\lambda_{ij}=0$ for $i\neq j$), population
of species \textit{i} eventually saturates at its carrying capacity $N^*_i
\equiv \mu_i/\lambda_{ii,}$  which is the stable fixed point of the logistic
growth process for species \textit{i}. We assume identical carrying capacities
  $N = N^*_1 = N^*_2$ throughout this paper.

In finite populations, microscopic rates in  (\ref{eqn:Birth}) and (\ref{eqn:Death})
define the Markov process for the stochastic dynamics in the number $N_i$
of  species \textit{i}. In the limit of large carrying capacity $1/N \ll
1$, the
discrete Master equation  for the joint probability distribution $P(N_1,N_2,t)$,
 associated with (\ref{eqn:Birth}) and (\ref{eqn:Death}), can now be approximated
by a continuous two-variable Fokker-Planck equation in the rescaled coordinates
$c_i \equiv
 N_i/N$:   
\begin{eqnarray}
\partial_t P(\boldsymbol{c},t) = \sum_{i=1}^2 \Big( &&-\partial_{c_i}[v_i(\boldsymbol{c})P(\boldsymbol{c},t)]+
\frac{1}{2N}\partial_{c_i}^2[D_i(\boldsymbol{c})P(\boldsymbol{c},t)]\Big),
\label{eqn: FP}
\end{eqnarray}  
where the deterministic drift and $N$-independent diffusion coefficients
read
\begin{eqnarray}
v_1(\boldsymbol{c})=\mu_1
c_1(1-c_1-c_2) +\mu_1
 \beta_1 c_1 c_2,\label{eqn: drift1}
\\
v_2(\boldsymbol{c})=\mu_2
c_2(1-c_1-c_2) +\mu_2 \beta_2 c_1 c_2,\label{eqn: drift2}
\\
D_1(\boldsymbol{c})= 
\mu_1 c_1(1+c_1+c_2) -\mu_1
 \beta_1 c_1 c_2 ,\label{eqn: diff1}
\\
D_2(\boldsymbol{c})= 
\mu_2 c_2(1+c_1+c_2) -\mu_2 \beta_2 c_1 c_2 ,\label{eqn: diff2}
\end{eqnarray}     
and  $\beta_1 \equiv 1-
 \Big(\frac{\lambda_{12}}{\lambda_{22}}\Big)\Big(\frac{\mu_2}{\mu_1}\Big)$
  and $\beta_2 \equiv 1-  \Big(\frac{\lambda_{21}}{\lambda_{11}}\Big)\Big(\frac{\mu_1}{\mu_2}\Big)$
are the rescaled parameters \cite{Chotibut:2015aa,Pigolotti:2013le}. The
inverse of the carrying capacity $1/N$ controls the relative strength of
deterministic    to fluctuating dynamics such that when $N \rightarrow \infty$
the dynamics is entirely deterministic and given by the coupled dynamical
equations: 
$dc_i/dt= v_i(\boldsymbol c).$

When $\beta_1 = \beta_2 = 0$  and when both species grow at the same rate
under dilute conditions ($\mu = \mu_1 = \mu_2 $), the stochastic dynamics
associated with the Fokker-Planck equation (\ref{eqn: FP}) describes neutral
evolution  with genetic drift without fixing the population size variable
$c_T \equiv c_1 + c_2$; the population size now fluctuates around the equilibrium
size at $N$ ($c_T = 1)$ at long times  \cite{Chotibut:2015aa,Pigolotti:2013le,Constable:2015aa}.
 This can be seen by  prescribing the Ito stochastic differential equations
associated with the Fokker-Planck equation (\ref{eqn: FP})  to obtain
coupled Langevin's dynamics in
the frequency   $f \equiv c_1/(c_1 + c_2)$  and the population size variable
 $c_T = c_1 + c_2$ \cite{Chotibut:2015aa}:
\begin{align}
\frac{df}{dt} &=\sqrt{\frac{\mu D_g(f)}{N}\left(\frac{1+c_{T}}{c_T}\right)}\Gamma_f(t),\label{eqn:dfdt_s0}\\
\frac{dc_T}{dt} &= \mu v_{G}(c_T)+\sqrt{\frac{\mu c_T(1+c_T)}{N}}\Gamma_{c_T}(t),\label{eqn:dcTdt_s0}
\end{align}  
where $\Gamma_i(t)$ is a  Gaussian white noise with $\langle\Gamma_i(t)\Gamma_j(t')
\rangle = \delta_{ij}\delta(t-t')$ and $\langle \Gamma_i(t) \rangle =
0$,  $D_g(f) =  f(1-f)$   
is  the frequency-dependent  genetic drift
coefficient, $v_G(c_T) \equiv c_T(1-c_T) $ is the logistic growth function.
Eq. (\ref{eqn:dcTdt_s0}) reveals that the population size variable undergoes
$f$-independent stochastic logistic growth dynamics   such that, at long
times, slow fluctuations with variance $1/N $ around
 the equilibrium  at $c_T=1$ take over (the equilibrium population size is
$N$). Near $c_T=1$, the  frequency dynamics of Eq. (\ref{eqn:dfdt_s0}) resembles
neutral evolution with genetic drift of Eq. (\ref{eqn: ito_rep_drift}). In
this case, the numerical
fixation probability $u(f)$ and the mean fixation times $\tau(f)$
starting along the equilibrium line $c_T =c_{1} + c_{2} = 1$ obtained via
the Gillespie algorithm show
excellent agreement with the standard population genetics results of  Eqs.
(\ref{eqn:fixprob_neutral})
and (\ref{eqn:MFT_neutral}) \cite{Pigolotti:2013le,Constable:2015aa}. Thus,
the competitive Lotka-Volterra model generalizes neutral evolution with genetic
drift to include independent population size fluctuations around the equilibrium
size without changing the essential results of neutral evolution, provided
$ \beta_1 = \beta_2 =0$ and $\mu_1 = \mu_2.$

\section{Fluctuation-induced Selection in quasi-neutral evolution}
We now discuss the scenario such that  $\beta_1 = \beta_2 = 0 $ but  $\mu_1/\mu_2
\neq
 1.$ As opposed to when $\mu_1/\mu_2 = 1$ in the previous section, the two
competing species no longer grow at an equal rate at low population densities.
Consequently, the mean frequency of each species is not fixed as the population
size grows up from small values and equilibrates at $c_T = 1$. This can be
seen as follows: In
the limit  $N \rightarrow \infty$, the dynamics are deterministic and are
given by 
\begin{align}
\frac{dc_1}{dt} &=\mu_1c_1(1-c_1-c_2), \label{Eqn:dc1dt}\\
\frac{dc_2}{dt} &= \mu_2 c_2 (1-c_1-c_2) \label{Eqn:dc2dt}. 
\end{align} 
The overall population size variable $c_T = c_1 + c_2$ grows and equilibrates
at
  $c_T = 1$ according to $dc_T/dt = (\mu_1 c_1 + \mu_2 c_2)(1-c_T).$ Moreover,
every point on the line $c_T =c_{1}+c_2 = 1$ is  a fixed point
(thus defining a fixed line). This limit defines neutral evolution
at the equilibrium population size since the frequency $f$ at the equilibrium
size is unchanged. Away from the equilibrium size, however, as the population
size
saturates, $c_1(t)$ and $c_2(t)$ change to conserve
the variable 
\begin{align}
\begin{split}
\rho &\equiv c_{2}(t) / c_1(t)^{(\mu_2/\mu_1)}, 
\label{eqn: Rho_Defn}
\end{split}
\end{align}  
since $d\rho/d t=0$  follows from Eqs. (\ref{Eqn:dc1dt})
and (\ref{Eqn:dc2dt}). Upon defining the \textit{selective
advantage in the dilute limit} (selective advantage near the origin) $s_o$
as $(1+s_o)
\equiv \mu_1/\mu_2$,  we can rewrite the conserved variable $\rho$  in terms
of  $f$ and $c_T $ as
\begin{equation}
\rho= c_T(t)^{s_{o}/(1+s_{o})} [1-f(t)]/f(t)^{1/(1+s_{o})}.\label{Eqn: Rho_f}
\end{equation}
 Eq. (\ref{Eqn: Rho_f}) and the conservation of $\rho$ imply that
the frequency of a species with a selective advantage in the dilute limit
increases (decreases) as $c_T(t)$ grows from $c_T(t)<1$ (declines from $c_T(t)>1$)
to  $c_T = 1$.
 This competition scenario with selective advantage away from, but neutral
at, the equilibrium   size (\textit{quasi}-neutral evolution) is illustrated
by the bent deterministic trajectories  that intersect the fixed line $c_T
= 1$ in Fig. \ref{fig: fluc-induced_drift}(b); the deterministic trajectory
is bent toward the axis of the species with a selective advantage in the
dilute limit. If both
species are also neutral in the dilute limit ($s_o = 0$), the relative frequency
is
conserved and the deterministic trajectory leading to the equilibrium fixed
line $c_T
= 1$ is a straight trajectory of fixed $f$; see Fig. \ref{fig: fluc-induced_drift}(a).

We now study the interesting limit of  finite populations,  in which, as
a result of the feedback
between $f$ and $c_T$,  fluctuation-induced selection at the equilibrium
population size emerges, even though the competing species are completely
neutral
at the equilibrium  size.   Without loss of generality, we assume that species
1 has a selective advantage in the dilute limit  ($1+s_o = \mu_1/\mu_2
>1$), and define $\mu \equiv \mu_2$ for brevity. Following Appendix A. of
Ref.
\cite{Chotibut:2015aa} and  absorbing $\mu$ into the unit of time, we find
the coupled stochastic
dynamics for $f$ and $c_T$ near $c_T=1$ in the limit $1/N \ll 1:$  
\begin{align}
\frac{df}{dt} = \ &v_{R}(f,c_{T}) + \sqrt{\frac{D_{g}(f)}{N}\left( \frac{1+c_T}{c_T}
\right)\Big(1+s_{o}(1-f)\Big)}\Gamma_f(t), \label{eqn:dfdt_s}\\
\frac{dc_T}{dt} = \ &(1+s_{o}f)v_{G}(c_T) +\sqrt{\frac{c_{T}(1+c_T)}{N}(1+s_{o}f)}\Gamma_{c_T}(t),\label{eqn:dcTdt_s}
\end{align}
where $v_R(f,c_T) =s_{o}f(1-f)   \left[(1-c_T) - \frac{1}{N}\left(\frac{1+c_T}{c_T}
\right) \right] $ is the  deterministic drift due to the selective advantage
near the origin $s_o$ and $v_G(c_T) = c_T(1-c_T)$ is the usual logistic
growth of population size.
Here, $\Gamma_f(t)$ and $\Gamma_{c_T}(t)$ are uncorrelated Gaussian white
noise with zero means and $\langle \Gamma_{\alpha}(t) \Gamma_{\beta}(t)\rangle
= \delta_{\alpha \beta} \delta(t-t'),$ interpreted according to Ito''s prescription.
In the standard neutral evolution, when $s_{o}=0$, 
  Eqs. (\ref{eqn:dfdt_s}) and   (\ref{eqn:dcTdt_s}) reduce to the  Moran
model for neutral evolution with a fluctuating population size  given
by
Eqs. (\ref{eqn:dfdt_s0}) and (\ref{eqn:dcTdt_s0}). In quasi-neutral evolution,
when $s_{o} \neq 0,$ however, fluctuations
of population size  becomes  $f$-dependent with  variance proportional to
$(1+s_{o}f)/N,$
while $f$ acquires an intriguing
deterministic drift at   $c_T = 1$ of the form  $v_R(f,c_{T}=1) = -2s_{o}f(1-f)/N
$,  which actually favors the fixation of  the species with a selective\textit{
disadvantage} near the origin ($s_o < 0$). The presence of non-vanishing
deterministic drift  is in
striking contrast to the unbiased random walk behavior
of neutral evolution along the equilibrium line $c_T = 1$ displayed in Eq.
(\ref{eqn:dfdt_s0}).
For the  generalization of   Eqs. (\ref{eqn:dfdt_s}) and  (\ref{eqn:dcTdt_s})
that reveals the role
of   a non-vanishing selection  in other non-neutral
scenarios, such as  mutualism,
see \cite{Chotibut:2015aa}. As this paper was nearing completion,
we  learned of related work in the context of public goods game by Constable
et al. \cite{Constable:2016aa}, who
studied the effect of two opposing selections: non-vanishing deterministic
selection that favors one species and the fluctuations-induced selection
that favors the other species. Such a scenario with two opposing selection pressures
also arose in the competitive Lotka-Volterra model studied in Ref. \cite{Chotibut:2015aa}, when $\beta_1$ and $\beta_2$
have opposite signs, $|\beta_1| \ll 1,$  $|\beta_2| \ll 1,$ and $1/N \ll
1$. In this  situation, fluctuation-induced selection
can reverse the direction of deterministic selection and alleviate the public
good dilemma of cooperation \cite{Constable:2016aa}. 

Eq. (\ref{eqn:dcTdt_s}) drives small excursions from $c_T = 1$ which changes
the dynamics of $f.$ To understand in more detail how quasi-neutral evolution
with a selective advantage near
the origin $s_o$ differs
from the classic Moran model, we  seek  an effective dynamics of $f$ near
the fixed
equilibrium population size $c_T =1$. Several works addressed  similar problems
 in the context of evolution \cite{Parsons:2007aa,Parsons:2008aa}, ecology
\cite{Lin:2012wf}, and epidemiology \cite{Kogan:2014aa}, and deduced
an
effective dynamics at the equilibrium line $c_T = 1$,  using
asymptotic expansions in powers
of
$1/N$ \cite{Lin:2012wf,Kogan:2014aa}. A more  systematic framework for studying
an effective dynamics  for stochastic dynamical systems with timescale separations
is discussed in Ref.  \cite{parsons2015dimension}. However, here, we present
an alternative (and, for us, more intuitive) argument
based on adiabatic elimination of a fast variable which exploits an appropriate
choice
of coordinates.  As
we shall see in Sec. \ref{subsec: 4C(dimensional_red)}, this  choice of coordinates
allows the fate of competitions to be inferred for an \textit{arbitrary}
population size, rather than constraining the description to $c_T \approx
1,$ i.e., near the
 equilibrium size. With the effective dynamics in hand,  we then  calculate
the fixation
probability as well
as the mean fixation time and verify the results with numerical simulations.
Our stochastic
simulations  employ the Gillespie algorithm to efficiently simulate the
discrete Master equation associated with 
 the microscopic rates (\ref{eqn:Birth}) and (\ref{eqn:Death}). The simulated
fixation probabilities and the mean fixation times for each
initial condition
are constructed from $10^4$ realizations of fixation events.

\subsection{A Naive Approximation}
\label{subsec: 4A(naive)}
\begin{figure}[t!]
\subfigure{\includegraphics[width=0.55\linewidth]{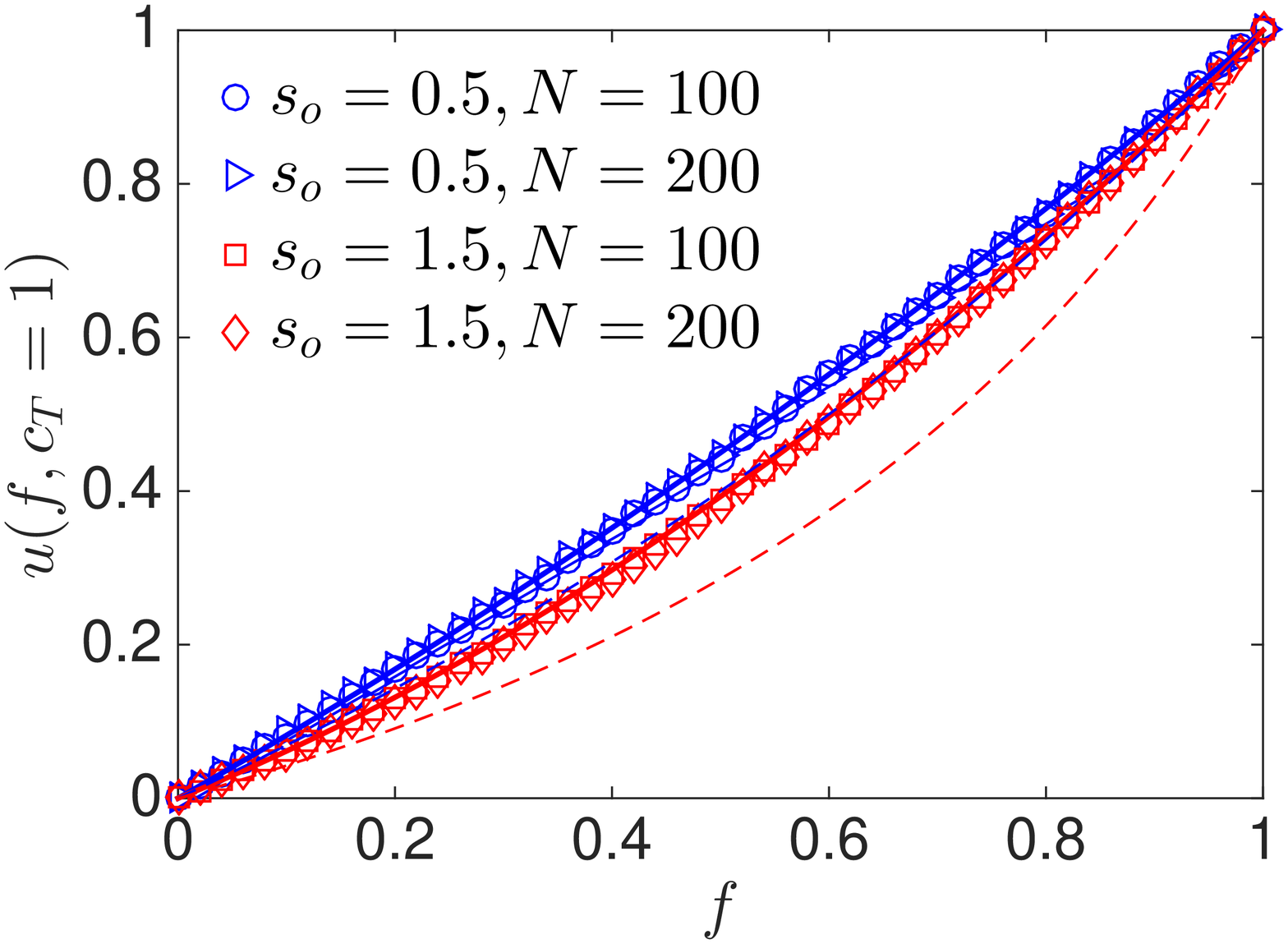}}
\subfigure{\includegraphics[width=0.55\linewidth]{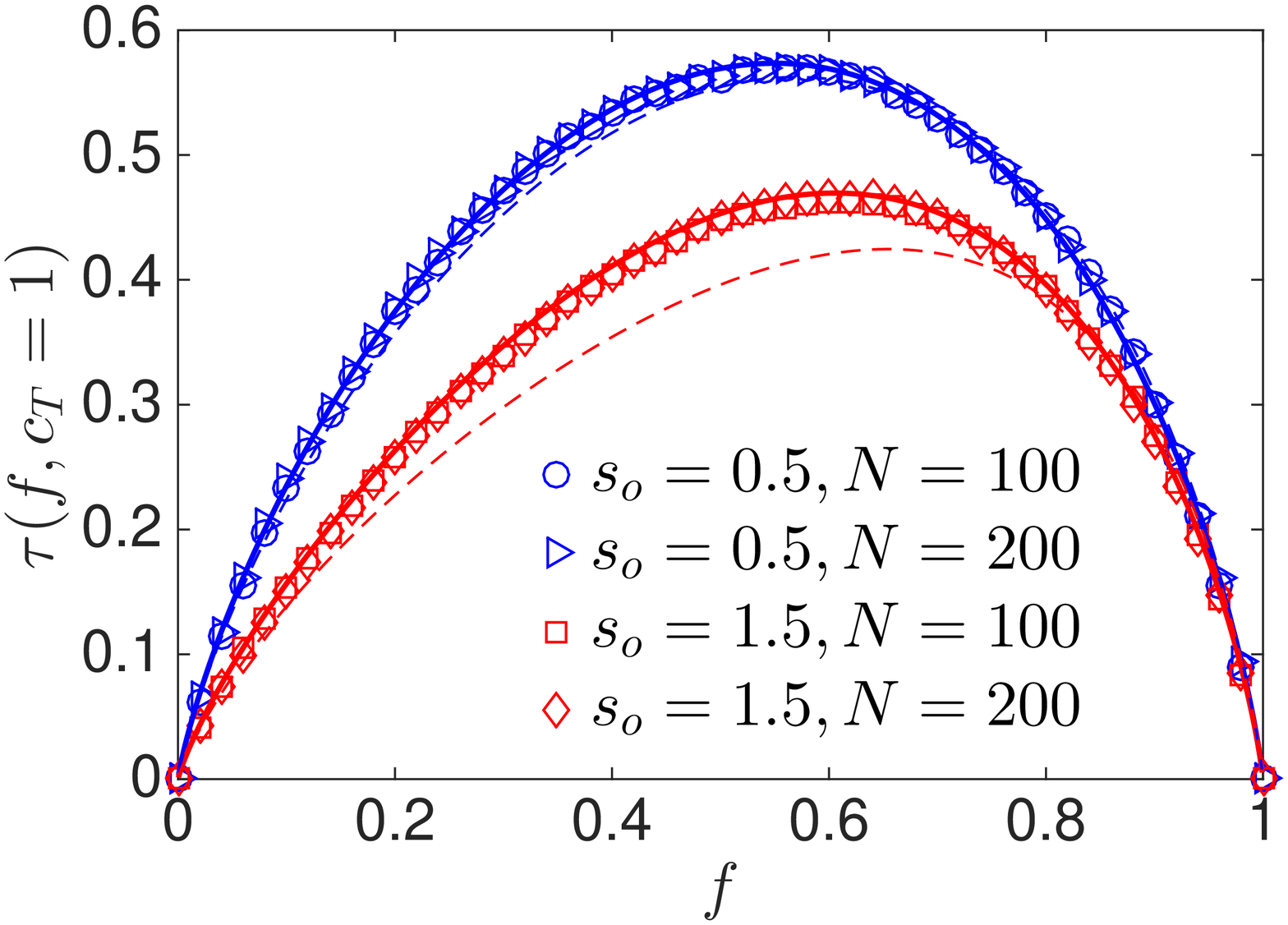}}
\caption{\label{fig: FP_MFT_quasi-neutral_ct1}  (Color Online) The   fixation
probability (top) and the mean fixation
time
in the units of $\mu/ N$  (bottom) as a function
of the initial frequency  $f$   with the initial overall population size
along the fixed line
$c_T
=1$. Predictions
from adiabatic elimination of a fast variable shown in solid lines are in
excellent agreement  with  simulations shown in symbols, while dashed lines
are predictions from our ``naive approximation," which exaggerates the deviations
from the classical results of Eqs. (\ref{eqn:fixprob_neutral}) and (\ref{eqn:MFT_neutral}).
The fixation probability
is
$N$-independent while the mean fixation time scales linearly with $N$, which
are also features
of unbiased random walk (genetic drift.) Population size fluctuations,
however,  induce selection  that disfavors  a species that grows faster
near the origin $(s_{o} > 0),$ resulting
in a decline in the fixation probability as well as a reduced mean fixation
time.} 
\end{figure}

\begin{figure}[t!]
\centering
\includegraphics[scale =0.4]{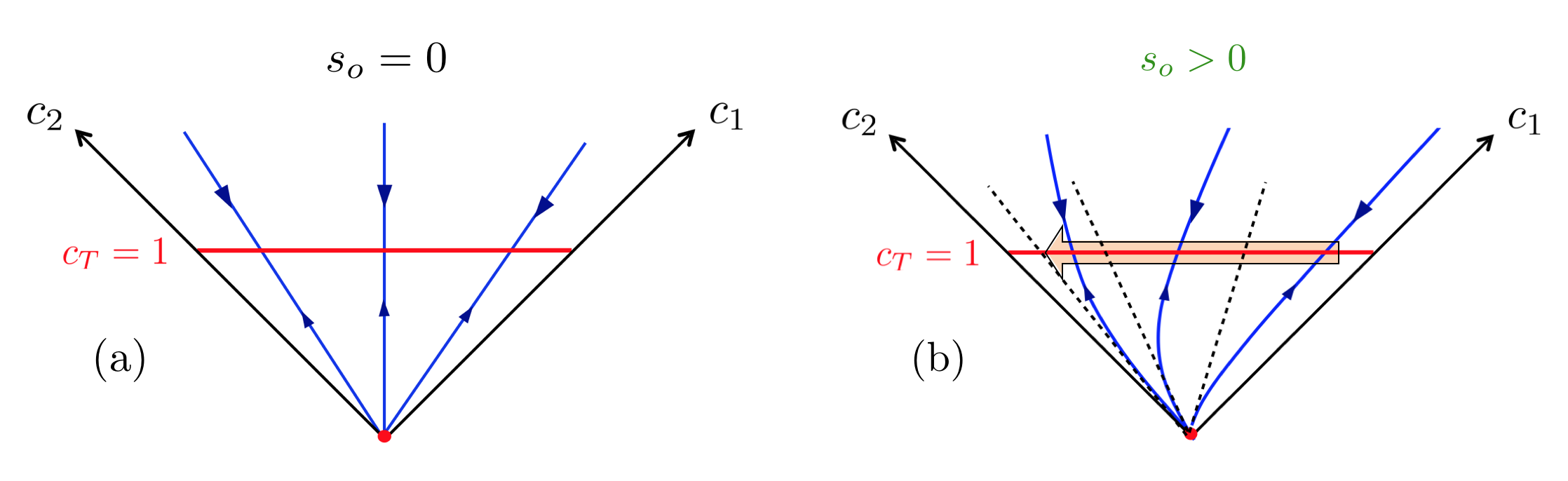}
\caption{\label{fig: fluc-induced_drift} 
(Color Online) Schematic phase portraits of deterministic neutral evolution
in (a) (left) and
quasi-neutral evolution in (b) (right) where species 1 is assumed to have
a selective
advantage near the origin ($s_o > 0$).  Blue curves represent deterministic
trajectories of dynamical systems in Eqs. (\ref{Eqn:dc1dt})-(\ref{Eqn:dc2dt})
that eventually reach the red fixed line $c_T = 1.$ Different curves correspond
to different values of the deterministically conserved variable $\rho$  of
Eq. (\ref{eqn: Rho_Defn}). For neutral evolution, the trajectories toward
the equilibrium population size $c_T =1$  are straight lines that fix the
fraction $f,$ while  the trajectories in quasi-neutral evolution bend toward
the axis of the species that grow faster in the dilute population limit (species
1 in this figure). In finite populations, a combination of
slow population
size fluctuations and  fast relaxation toward
 $c_T = 1$ along a warped
trajectory $c_2 = (\rho c_1)^{1+s_o}$ of conserved $\rho$, depicted by the
blue curves, generates an effective
selection at $c_T \approx 1,$ depicted by the faint orange arrow on the right.
Remarkably, fluctuation-induced
selection \textit{disfavors} fixation of species with a  selective advantage
near the origin (the faint orange arrow points away from the axis of the
species with
$s_o >0$).  
 The effective stochastic dynamics of $f$ when $c_T \approx 1$ after adiabatic
elimination 
 of the fast variable $c_T$ is given by Eq. (\ref{eqn:dfdt_correct}). 
 The effective selection term $\tilde v(f)$ given by Eq. (\ref{eqn: effective_drift})
differs from the naive approximation  $v_R(f,c_T=1)$
 in Eq. (\ref{eqn:dfdt_naive}) by a factor of $2(1+s_of)^2/(1+s_o)$,  
 leading to the improved agreement between simulation and theory when $\mathcal{O}(s_o)
\sim 1$ in Fig. \ref{fig: FP_MFT_quasi-neutral_ct1}.}
\end{figure}

In the limit $1/N \ll 1,$ one strategy to close   Eq. (\ref{eqn:dfdt_s})
for $f$
is to substituting $c_T = 1 $  and ignore weak population size fluctuations
of order
$1/N.$ This naive
approximation yields
\begin{equation}
\frac{df}{dt} = -\frac{2s_{o}}{N}f(1-f)+ \sqrt{\frac{2D_{g}(f)}{N}\Big(1+s_{o}(1-f)\Big)}\Gamma_f(t).
\label{eqn:dfdt_naive}
\end{equation}
   Upon solving the associated backward Kolmogorov equations (similar to
solving
Eqs. (\ref{eqn: bke_u}) - (\ref{eqn:
bke_tau}) associated with the stochastic differential equation (\ref{eqn:
ito_rep_drift})),  we determine $s_o$-dependent corrections to the fixation
probability
and the mean fixation time, 
\begin{align}
u(f) &=\frac{u_{neutral}(f)}{1+s_{o}(1-f)}, \label{eqn: FixedProb_Naive}
\\
\tau(f) &= \frac{\tau_{neutral}(f)}{1+s_{o}(1-f)},\label{eqn: MFT_Naive}
\end{align}
where $u_{neutral}(f)$ and $\tau_{neutral}(f)$ are given by Eqs. (\ref{eqn:fixprob_neutral})
and  (\ref{eqn:MFT_neutral}).
Although at $s_{o}=0$ we recover the results of the Moran model, dashed lines
in Fig. \ref{fig: FP_MFT_quasi-neutral_ct1} show that these  approximations
are poor when $s_o = 0.5$ and $s_o = 1.5$. These errors stem from  neglecting
overall population
size fluctuations of the term $s_{o}f(1-f)(1-c_T),  $  which (as we shall
see) actually contribute
a deterministic
drift of order $s_{o}/N $ comparable to the term kept in Eq. (\ref{eqn:dfdt_naive}).

\subsection{Effective Evolutionary Dynamics Near The Equilibrium Population
Size}
\label{sec: EffectiveEvo}

We now employ adiabatic elimination of a fast variable to deduce an effective
evolutionary dynamics for an approximately fixed  population size close
to $N.$ Motivated by the approximate conservation of the composite variable
$\rho$
neglecting number fluctuations (see Eq. (\ref{eqn: Rho_Defn})),
we calculate the stochastic dynamics of $\rho$ from Ito's change of variable
formula \cite{Van-Kampen:1992tk,Gardiner:1985qr}, and find     
\begin{equation}
\frac{d\rho}{dt}=\frac{v_\rho(f,c_T)}{N}+ \sqrt{\frac{2D_\rho(f,c_T)}{N}}\Gamma_\rho(t),\label{eqn:rho_stochastic}
\end{equation} 
where 
\begin{eqnarray}
v_\rho(f,c_T) = \frac{1}{2} \left(\frac{2+s_{o}}{1+s_{o}}\right)\left(\frac{1-f}{f}\right)
\left(\frac{1+c_T}{c_T}\right) \left(\frac{c_T^{s_o}}{f}\right)^{1/(1+s_o)},
\label{eqn: v_rho}
\end{eqnarray}
\begin{eqnarray}
D_\rho(f,c_T)=\frac{1}{2}\left (\frac{1+s_{o}f}{1+s_{o}}\right)\left(\frac{1-f}{f}\right)\left(\frac{1+c_T}{c_T}\right)
\left(\frac{c_T^{s_o}}{f}\right)^{2/(1+s_{o})},
\label{eqn: diff_rho}
\end{eqnarray}
and $\langle \Gamma_\rho(t) \Gamma_\rho(t') \rangle = \delta(t-t').$
 Eqs. (\ref{eqn:rho_stochastic})-(\ref{eqn: diff_rho}) reveal that   $\rho$
varies  on a slow timescale
of order $1/N\ll1$ everywhere in our domain of interest. On the other hand,
the dynamics
of the overall population size  given by  Eq. (\ref{eqn:dcTdt_s}) exhibits
a fast relaxation
toward $c_T \approx 1, $ after which slow fluctuations of order $1/N$ take
over.
Since $\rho=(c_{2}/c_{1})^{(\mu_2/\mu_1)}$ and $c_T = c_1 + c_2$ together
completely specify the state of the system,
the dynamics of the system starts with
a rapid quasi-deterministic relaxation toward $c_T \approx 1$  along a trajectory
of
fixed
$\rho;$ then the slow residual dynamics of $\rho$ takes over. The slow dynamics
of the  coordinate $\rho$ generates an effective dynamics
of $f$ when $c_T \approx 1$. Fig. \ref{fig: fluc-induced_drift}(a) 
depicts the fluctuation-induced selection emerging from the slow stochastic
dynamics
of $\rho$ near $c_T=1$.   
 
To explicitly eliminate the fast variable, we  integrate out $c_T$ in the
joint
probability distribution of  $c_T$ and $\rho$
at time $t,$ 
$P(c_T,\rho,t), $ and obtain  the marginal
probability
distribution $\tilde P(\rho,t) \equiv \int P(c_T, \rho,t) dc_T.$ 
The Fokker-Planck equation for the marginal probability distribution dictates
the effective dynamics of the remaining slow variable $\rho.$ Motivated by
the separation of timescales, we factorize  $P(c_T, \rho,t) = P_{st}(c_T)P_\rho(\rho,t),$
assuming $c_T$ rapidly relaxes to $c_T=1$ and
forms a quasi-stationary distribution $P_{st}(c_T)$ before $\rho$ varies
significantly.
In other words, $c_T$ is slaved  to $\rho$ \cite{Gardiner:1985qr}. Upon substituting
this factorization into the Fokker-Planck
equation associated
with   Eqs. (\ref{eqn:dcTdt_s}) and (\ref{eqn:rho_stochastic}), we find 
\begin{eqnarray*}
 \partial_tP(c_T,\rho,t) &&= -\vec\nabla \cdot\vec J(c_T,\rho,t)\\
 &&=-\bigg[\frac{1}{N}\partial_{\rho}v_{\rho}(f,c_T)P_\rho(\rho,t)-\frac{1}{N}\partial^2_{\rho}D_{\rho}(f,c_T)P_\rho(\rho,t)\bigg]P_{st}(c_{T}),
\end{eqnarray*}
where the probabilistic current in the $c_T$ direction vanishes by the
assumption of stationarity. Integrating out $c_T$ then leads to
\begin{eqnarray}
\partial_t \tilde P(\rho,t) = -\frac{1}{N}\partial_\rho v_{\rho}(f, \langle
 c_T\rangle)\tilde
P(\rho,t) + \frac{1}{N}\partial^2_\rho D_{\rho}(f, \langle c_T \rangle)\tilde
P(\rho,t), \label{eqn:F-P_rho}
\end{eqnarray}
where $\langle . \rangle$ denotes an expectation value. The effective Langevin
dynamics
associated with Eq. (\ref{eqn:F-P_rho})  is precisely 
Eq. (\ref{eqn:rho_stochastic}) with the substitution $c_T = \langle c_T
\rangle$, which is here simply the equilibrium population size $\langle
c_T \rangle
=1.$ 

We can now determine the effective evolutionary dynamics when $c_T \approx
1$
by substituting
$\rho(f)$  for $c_T \approx 1$, i.e.  (using Eq. (\ref{Eqn: Rho_f})) we have
$\rho
= (1-f)/f^{1/(1+s_{o})}.$ The Fokker-Planck equation for $\rho$
can be converted to the Fokker-Planck equation for $f$ along the line $c_T
= 1$ via the chain rule  $d/d\rho =-[(1+s_o)f(1-f)/\big((1+s_{o}f)\rho\big)]d/df.$
The calculation is more easily carried out using the 
backward Kolmogorov equation, since derivatives only act on the probability
distribution. A straightforward calculation leads to an effective Fokker-Planck
equation of
$f$ for $c_T \approx 1$; namely,
\begin{equation}
\partial_t \tilde P(f,t) = -\partial_f \tilde v(f) \tilde P(f,t)
+ \frac{1}{N}\partial^2_f \tilde D(f)\tilde
P(f,t), \label{eqn: F-P_f_effective}
\end{equation}    
where 
\begin{equation}
\tilde v(f) =- \left( \frac{1}{N}\right) s_{o}(1+s_{o})
 \frac{f(1-f)}{(1+s_{o}f)^2},
 \label{eqn: effective_drift}
\end{equation}
and
\begin{equation}   
\tilde D(f) =D_{g}(f)\left( \frac{1+s_{o}}{1+s_{o}f}\right) 
\label{eqn: effective_diff}.
\end{equation}    
Hence, the effective dynamics of $f$  reads
\begin{equation}
\frac{df}{dt} = \tilde v(f)  + \sqrt{\frac{2\tilde D(f)}{N}}\Gamma_f(t),
\label{eqn:dfdt_correct}
\end{equation}
where $ \tilde v(f)$ is given by Eq. (\ref{eqn: effective_drift}) and describes
fluctuation-induced selection term (displayed as the faint orange arrow in
Fig. \ref{fig: fluc-induced_drift}(b),
and  $\tilde
D(f)$ is the effective genetic drift coefficient given by Eq. (\ref{eqn:
effective_diff}). Eq. (\ref{eqn:dfdt_correct}) reduces
to the  Moran model for neutral evolution at $s_{o} = 0$. For $s_{o} \neq
0$, not
only does fluctuation-induced selection appear, but we also obtain an effective
genetic
drift that differs from the Wright-Fisher sampling by a frequency-dependent
factor $(1+s_{o})/(1+s_{o}f)$. 

The fixation probability and the mean fixation time with an initial condition
on the equilibrium line $c_T = 1$ now follow immediately from
  solving the Backward Kolmogorov equations associated with Eq.  (\ref{eqn:dfdt_correct}):
\begin{equation}
u(f) = \frac{(2+s_{o}f)}{(2+s_{o})}u_{neutral}(f),
\label{eqn: FixedPop_Correct}
\end{equation}
\begin{eqnarray}
\tau(f) = -\left(\frac{N}{\mu}\right)\Bigg[
\left(\frac{1+\frac{s_{o}f}{2}}{1+s_{o}}\right) f \ln f &&+ \left(\frac{1+\frac{s_{o}(1-f)}{2}}{1+s_{o}}\right)(1-f)\ln(1-f)\nonumber\\
&&+ \frac{s_{o}^2}{2(1+s_{o})(2+s_{o})}f(1-f)\Bigg].\label{eqn:MFT_correct}
\end{eqnarray}
Eqs. (\ref{eqn:dfdt_correct})-(\ref{eqn:MFT_correct}) are in
agreement with the results of Refs. \cite{Lin:2012wf,Parsons:2007aa,Parsons:2008aa}
 after an appropriate change of variable. At small
$s_{o}, $  both Eq. (\ref{eqn:
FixedPop_Correct}) and Eq. (\ref{eqn:
FixedProb_Naive}) give $u(f)=[1-s_{o}(1-f)]u_{neutral}(f)+\mathcal{O}(s_{o}^2)$
while both
Eq. (\ref{eqn:MFT_correct}) and Eq. (\ref{eqn: MFT_Naive}) give  $\tau(f)=[1-s_{o}(1-f)]\tau_{neutral}(f)+\mathcal{O}(s_{o}^{2}),$
reducing to the standard results of the Moran model when $s_{o} =0.$    
The differences appear only at $\mathcal{O}(s_o^2).$ Fig. \ref{fig: FP_MFT_quasi-neutral_ct1}
shows  the predictions from
Eq. (\ref{eqn: FixedPop_Correct}) and (\ref{eqn:MFT_correct})
are in excellent agreement with our stochastic  simulations.

Upon inoculating an equal mixture of each species and assuming species 1
has a selective advantage near the origin $s_o$, the fixation probability
of species 1 is
$1/4+1/(4+2s_{o})$ which monotonically decreases from $1/2$ when $s_{o }
= 0$
to $1/4$ as $s_o \rightarrow \infty$. Moreover,  
by defining $\tilde f$ such that the fixation probability $u(\tilde f) =
1/2, $ we find $\tilde f = (2+s_{o})/\big(2+\sqrt{4+2s_{o}(2+s_{o}})\big)$
  which rises monotonically
from $\tilde f = 1/2$ when $s_o = 0$ to $\sqrt{2}/2$ as $s_o \rightarrow
\infty$. Consequently, the faster growing species near the origin is only
more likely to survive provided the initial fraction is biassed in its favor,
$f \in [\sqrt{2}/2, 1] \approx [0.707,1]$ for $c_T = 1$, confirming that
population size fluctuations \textit{disfavor} the ultimate survival of a
species
with a selective advantage near the origin.

\subsection{Dimensional Reduction: the Fixation Probability and the Mean
Fixation Time}
\label{subsec: 4C(dimensional_red)}
\begin{figure*}[ht!]
\includegraphics[width = \textwidth]{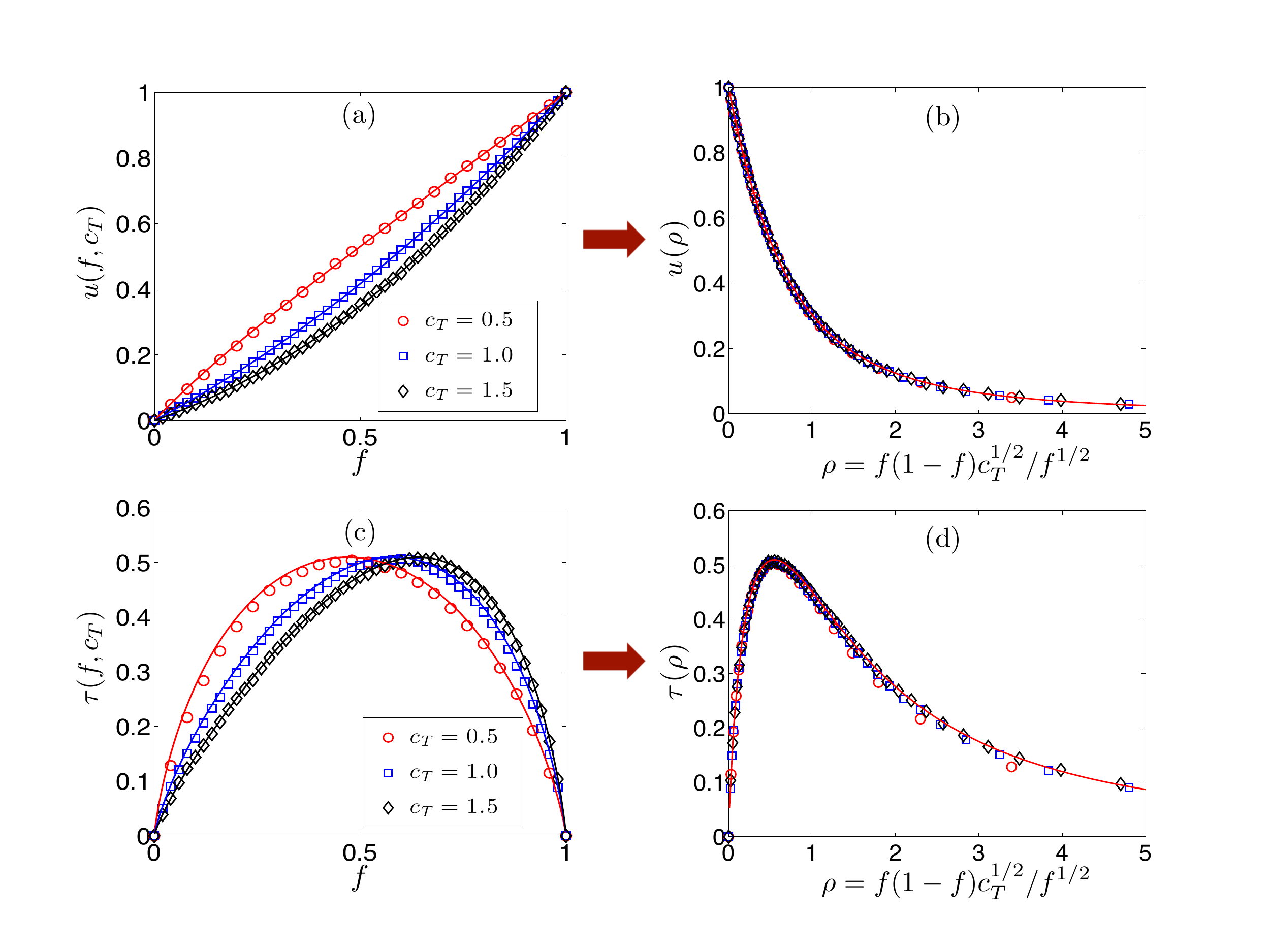}
\caption{\label{fig: QuasiNeutral_FitTheory} (Color Online) Dimensional reduction
from 2 variables to 1 variable by adiabatic elimination of
the fast population size variable
 for $s_{o} = 1$ and $N = 100$. In all figures, solid lines are analytical
predictions constructed in Sec. \ref{sec: EffectiveEvo}
while symbols are simulation results.  (a) and (c) show the fixation probability
and
the mean fixation time  as a function of initial frequency $f$ in a dilute
($c_T =0.5$), optimal ($c_T
= 1$), and overcrowded ($c_T = 1.5$) initial population size.  
When replotted against the slow variable $\rho,$ (b) and (d) show data collapse
of the fixation probability and the mean fixation time onto $u_{s_o=1}(\rho)$
and $\tau_{s_o=1}(\rho)$.
(a) demonstrates that population
size degree of freedom plays a crucial role in determining the fate of
competition; a wise strategy for the  species with a selective advantage
near the origin
is  to start with a dilute population size. On the other hand, a species
with a selective disadvantage near the origin  is better off starting
in an overcrowded
population size.}
\end{figure*}
Since the dynamics also contains the overall population size degree of freedom,
the initial frequency $f^0$ and the initial population size $c_T^0$ will
both in general 
enter the fixation probability $u(f^0, c_T^0)$ and the mean fixation time
$\tau(f^0, c_T^0)$.
To keep the notation simple, we now continue with the practice of setting
$f\equiv f^0$ and $c_T \equiv c_T^0$. Separation of dynamical timescales,
in fact, implies the fixation probability
and the mean fixation time are universal functions of the slow variable $\rho
= (1-f)f^{-1/(1+s_{o})}c_T^{s_{o}/(1+s_{o})}$
 , provided $1/N \ll 1.$  In other words, $u(f,c_T)=u(f',c_{T}')=u(\rho)$
and $\tau(f,c_T)=\tau(f',c'_T)=\tau(\rho) $ if $\rho(f,c_T)=\rho(f',c'_T)=\rho$.
These simplifications arise from a rapid quasi-deterministic
relaxation of the population size, with $\rho$ fixed,  toward the line $c_T
= 1$, after which  the slow stochastic dynamics of $\rho$ dictates
the  outcome.     

In principle, $u(\rho)$ and $\tau(\rho)$ follow from  rewriting
$f$ as a function
of $\rho$ in Eqs. (\ref{eqn:
FixedPop_Correct})
and (\ref{eqn:MFT_correct}). For  an arbitrary $s_{o},$\ however, $f$ cannot
easily be expressed as a function of $\rho$ at $c_T = 1$ because
they
are related by an $s_o$-dependent transcendental equation 
\begin{equation}
\rho(f,c_T=1) = (1-f)f^{-1/(1+s_{o})}. \label{eqn: rho_ct1_f}
\end{equation}
One can  nevertheless extract   $u(\rho)$ and $\tau(\rho)$ from   Eqs. (\ref{eqn:
FixedPop_Correct})
and (\ref{eqn:MFT_correct})
by numerically solving Eq. (\ref{eqn: rho_ct1_f}). 

Consider the particularly simple case  $_{}s_{o}=1,$ where
the physically relevant closed-form solution
 associated with  Eq. (\ref{eqn:
rho_ct1_f}) is $ f(\rho) = 1+\frac{1}{2}\left( \rho^2 - \rho\sqrt{\rho^2+4}
\right).$ Substituting
$f(\rho)$
into  Eqs. (\ref{eqn:
FixedPop_Correct})
and (\ref{eqn:MFT_correct})
now yields analytical results of  $u_{s_o=1}(\rho)$ and $\tau_{s_o=1}(\rho).
$
It is also possible to reconstruct  $u_{s_o=1}(f,c_T)$
and $\tau_{s_o=1}(f,c_T)$ for arbitrary $f$ and $c_T$ from    $u_{s_o=1}(\rho)$
and $\tau_{s_o=1}(\rho)$ by a direct substitution $\rho=
(1-f)f^{-1/2}c_T^{1/2}$ .
To test these predictions, we  simulated $10^4$ fixation events per each
initial
condition, with $N = 100$ and with  $c_T = 0.5, \ c_T = 1, \ c_T = 1.5$ representing
dilute,
optimal, and overcrowded initial population sizes with $s_{o}=1.$  
 Figs. \ref{fig: QuasiNeutral_FitTheory}(a)
and \ref{fig:
QuasiNeutral_FitTheory}(c)  show excellent agreement between $u_{s_o=1}(f,c_T)$
as well as $\tau_{s_o=1}(f,c_T)$ and the simulations. Figs. \ref{fig:
QuasiNeutral_FitTheory}(b) and \ref{fig:
QuasiNeutral_FitTheory}(d) show data collapse of the fixation probability
and the mean fixation time onto   $u_{s_o=1}(\rho)$ and $\tau_{s_o=1}(\rho)$
constructed above. These results demonstrate
that population size degree of freedom can play
a crucial role in determining the results of competition, here
through the composite variable $\rho= (1-f)f^{-1/(1+s_{o})}c_T^{s_{o}/(1+s_{o})}$.

\section{Conclusion}
\label{sec:Conclusion}

We began by reviewing standard theory  of neutral evolution of well-mixed
systems with a fixed population size
in population genetics using the language of statistical
physics. A competitive Lotka-Volterra model that exhibits both neutral evolution
and independent fluctuations in the population size was introduced. Relaxing
the fixed population size assumption leads to  interesting fluctuation-induced
 phenomena, such that the feedback between evolutionary dynamics and population
size fluctuations induces a selective advantage for the species that grow
faster in the dilute population even when the two competing species are neutral
at the equilibrium size. In this situation, there is a natural separation
of timescales between the fast population size variable $c_T$ and the slow
composite variable         $\rho$ that depends on both the relative frequency
$f$ and the population size  $c_T.$ Because of this  separation of timescales,
the effective evolutionary dynamics near an equilibrium population size can
be
deduced by means of adiabatic elimination of a fast variable, which reveals
a fluctuation-induced selective advantage and unusual genetic drift of a
non-Wright-Fisher (or non-Moran) type. In addition, we found that the fixation
probability
and the mean fixation time are universal functions of the slow composite
variable $\rho,$  allowing the fate of competitions at an \textit{arbitrary}
initial population size to be deduced. Given a fixed initial frequency $f$,
a better strategy for the species that grows fast in the dilute limit
 to ultimately fix (i.e., take over the populations) is to begin with both
 populations dilute ($c_T < 1$), rather
than overcrowded populations ($c_T > 1$).  Unlike the generalization from
canonical ensemble to grand canonical ensemble in equilibrium statistical
mechanics,
replacing the population size by its average value does not yield the accurate
description  due to the intricate coupling between the frequency and the
population size. These findings indicate the importance
of the population size  variable in population genetics results for the fixation
probability and the fixation time. 

It is a pleasure to dedicate this paper to the memory of Leo Kadanoff.    One of us (drn) owes a particular debt to Leo, for a collaboration (J. V. Jos\'e et. al., Physical Review B16, 1217 (1977)) that provided an inspiring example of how to do theoretical physics early in his scientific career.
  
\begin{acknowledgments}
This work was supported in part by the National Science Foundation (NSF) through Grants No. DMR-1608501  and DMR-1306367 and by the Harvard Materials Research Science and Engineering Laboratory, through MRSEC Grant No. DMR-1420570. Portions of this research were conducted during a stay at the Center for Models of Life at the Niels Bohr Institute, the University of Copenhagen. Computations were performed on the Odyssey cluster supported by the FAS Division of Science Research Computing Group at Harvard University.
\end{acknowledgments}


\end{document}